\documentclass[]{article}
\usepackage{emulateapj}
\usepackage{graphicx}
\usepackage{psfig}

\def\msun{{\rm ~M}_{\odot}}

\begin{document}

\title{A Study of Compact Object Mergers as Short Gamma-ray Burst Progenitors}

 \author{Krzysztof Belczynski\altaffilmark{1,2}, Rosalba Perna\altaffilmark{3},
         Tomasz Bulik\altaffilmark{4,5}, Vassiliki Kalogera\altaffilmark{6},
         Natalia Ivanova\altaffilmark{7} and Donald Q. Lamb\altaffilmark{8}}

 \affil{
     $^{1}$ New Mexico State University, Dept. of Astronomy,
            1320 Frenger Mall, Las Cruces, NM 88003\\
     $^{2}$ Tombaugh Fellow\\
     $^{3}$ JILA \& Dept. of Astrophysical \& Planetary Sciences, University of
            Colorado, 440 UCB, Boulder, CO 80309\\
     $^{4}$ Astronomical Observatory, Warsaw University, Al Ujazdowskie 4, 00478 Warsaw
            Poland;\\
     $^{5}$ Nicolaus Copernicus Astronomical Center, Bartycka 18, 00716 Warsaw, Poland\\
     $^{6}$ Northwestern University, Dept. of Physics \& Astronomy,
            2145 Sheridan Rd., Evanston, IL 60208\\
     $^{7}$ CITA, University of Toronto, 60 St. George, Toronto, ON M5S 3H8,
            Canada\\
     $^{8}$ Dept. of Astronomy \& Astrophysics, University of Chicago, 5640
            S. Ellis Av., Chicago, IL 60637\\
     kbelczyn@nmsu.edu,rosalba@jilau1.Colorado.EDU,bulik@camk.edu.pl,
     vicky@northwestern.edu,nata@cita.utoronto.ca,lamb@oddjob.uchicago.edu}

 \begin{abstract} 
We present a theoretical study of double compact objects as potential
short/hard gamma-ray burst (GRB) progenitors. An updated population
synthesis code {\tt StarTrack} is used to calculate properties of
double neutron stars and black-hole neutron star binaries. We obtain
their formation rates, estimate merger times and finally predict their
most likely merger locations and afterglow properties for different
types of host galaxies. Our results serve for a direct comparison with
the recent {\em HETE-II} and {\em SWIFT} observations of several short
bursts, for which afterglows and host galaxies were detected. We also
discuss the possible constraints these observations put on the
evolutionary models of double compact object formation. We emphasize
that our double compact object models can successfully reproduce at
the same time short GRBs within both young, star-forming galaxies
(e.g., GRB~050709 and GRB~051221A), as well as within old, 
elliptical hosts (e.g.,  GRB~050724 and probably GRB~050509B).
\end{abstract}


\section{Introduction}

The recent detections of afterglows for short/hard GRBs have made
possible a breakthrough in the study of this class of bursts. {\em
Swift} (Gehrels et al. 2005; Barthelmy et al. 2005) and {\em HETE-II} 
(Villasenor et al. 2005) observations led to precise localizations of
several short GRBs, and subsequent follow-up optical observations
allowed for tentative connections with their host galaxies and a
measurement of their redshifts.  GRB~050509B was found to lie in the
vicinity of a large elliptical galaxy, with no current star formation
and at a redshift of 0.225 (Gehrels et al. 2005).  GRB~050709 was found
in the outskirts of a dwarf irregular galaxy with ongoing star
formation at redshift 0.1606 (Hjorth et al. 2005; Fox et al. 2005;
Covino et al. 2006), as was GRB~051221A (Soderberg and Berger 2005,
Berger and Soderberg 2005) at redshift 0.5465. GRB~050724 was found
within a small elliptical galaxy with no current star formation at a
redshift of 0.258 (Berger et al. 2005a).  GRB~050813 was found close to
a cluster of galaxies at redshift 1.7-1.9 (Berger 2005b), but no host
galaxy has been proposed.

Compact object mergers from double neutron star (NS-NS) and black hole
neutron star (BH-NS) binaries have been proposed for the first time by 
Paczynski (1986) and then further discussed as the central engines of 
short GRBs by number of authors (e.g., Eichler et al. 1989; Narayan,
Paczynski \& Piran 1992). Observational constraints on these
populations may be obtained only for NS-NS systems since only such
binaries are currently observed as binary pulsars. Merger rates derived
from the observed sample of a handful of Galactic
relativistic NS-NS binaries have been presented by Kalogera et al.\
(2004 and references therein). Double compact objects with both neutron
stars and black holes can be studied via population synthesis methods
(e.g., Lipunov, Postnov \& Prokhorov 1997; Portegies Zwart \& Yungelson
1998; Belczynski, Kalogera \& Bulik 2002c) There were several early
population synthesis studies in the context of potential GRB
progenitors (e.g., Bloom, Sigurdsson \& Pols 1999; Belczynski \& Bulik
1999; Fryer, Woosley \& Hartmann 1999). In particular, it was found
that NS-NS and BH-NS mergers are expected to take place outside host
galaxies with long delay times.  However, it has only recently  
been recognized that the population of the double compact objects may
be more diverse than was previously believed (Belczynski \&
Kalogera 2001; Belczynski, Bulik \& Rudak 2002b; Perna \& Belczynski
2002; Belczynski, Bulik \& Kalogera 2002a). In addition to classical
well-recognized channels, these newly recognized formation channels
lead to the formation of tighter double compact objects, with short
lifetimes and therefore possible prompt mergers within hosts.  The new
formation scenarios were independently confirmed by detailed
evolutionary calculations (Ivanova et al. 2003; Dewi \& Pols 2003).

In this study we perform an updated analysis of double compact object
mergers using population synthesis methods. Over the last several
years and since our previous studies, the {\tt StarTrack} population
synthesis code has undergone major revisions and updates, many of
which are guided and tested through comparisons with either
observations or detailed evolutionary calculations (see Belczynski et
al. 2005). Additionally, this new study is motivated by the recent
short GRB observations and their likely connection with double compact
object mergers. In \S\,2 we present the overview of our
claculations. In \S\,3 the double compact object formation, merger
rates, locations and their afterglow  properties are
presented. Finally, in \S\,4 we discuss our results in context of the
recent short-GRB observations.

\section{Binary Compact Object Models}

{\underline{\em Binary Population Synthesis.}}  The {\tt StarTrack}
population synthesis code was initially developed for the study of
double compact object mergers in the context of GRB progenitors
(Belczynski et al. 2002b) and gravitational-wave inspiral sources
(Belczynski et al. 2002c). In recent years {\tt StarTrack} has
undergone major updates and revisions in the physical treatment of
various binary evolution phases. The new version has already been
tested against observations and detailed evolutionary calculations
(Belczynski et al.\ 2005), and has been used in various applications
(e.g., Belczynski \& Taam 2004; Belczynski et al.\ 2004a; Belczynski,
Sadowski \& Rasio 2004b).  The most important updates for compact
object formation and evolution include: a full numerical approach to
binary evolution due to tidal interactions and coupling calibrated
using high mass X-ray binaries and open cluster observations, a
detailed treatment of mass transfer episodes fully calibrated against
detailed calculations with a stellar evolution code, updated stellar
winds for massive stars, and the latest determination of natal kick
velocity distribution for neutron stars (Hobbs et al.\ 2005). In
the helium star evolution, which is of a crucial importance for the
formation of new classes of double compact objects (e.g., Ivanova et
al.\ 2003), we have applied a conservative treatment matching closely
the results of detailed evolutionary calculations. The NS-NS
progenitors are followed and checked for any potential Roche lobe
overflow (RLOF). While in the mass transfer phase, systems are examined for
potential development of dynamical instability, in which case the systems
are evolved through a common envelope phase. 
We treat common envelope events through the energy formalism (Webbink 1984;
Belczynski et al. 2002), where the binding energy of the envelope 
is determined from the set of He star models calculated with the
detailed evolutionary code by Ivanova et al. (2003).
For some systems we observe, as before, extra orbital decay leading to
the formation of very tight short lived double compact object binaries.
However, since the progenitor evolution and the final RLOF episodes 
are now followed in much greater detail, we note significant
differences from our earlier studies. For a detailed description of the
revised code we refer the reader to Belczynski et al.\ (2005).

{\underline {\em Galaxy Potential Models.}}  To investigate the motion
of binary compact objects in their host galaxies and study the merger
locations we consider a set of typical gravitational potential models
for different types of galaxies. Our model spiral galaxy is identical
to the one used in Belczynski et al.\ (2002b).  It consists of a disk,
bulge, described by a Miyamaoto \& Nagai (1975) type potential and a
halo with the dark matter density of
$\rho=\rho_c[1+{r/r_c}^2]^{-1}$. We consider: a large spiral similar
to the Milky Way, with a disk and bulge mass of $10^{11} \msun$ and a
massive halo of $10^{12} \msun$ extending out to $100$kpc; and a small
spiral downscaled by a factor of $10^3$ in mass and of $10$ in size
(constant density). We place the binaries on circular orbits in the
disk of a spiral galaxy. Binaries with the full range of delay times
(i.e., time from the formation of a double compact object binary until
the merger) are used here since spirals consists of both old and young
populations due to a roughly continuous star formation history.  We
also consider elliptical host galaxies. The model potential of an
elliptical galaxy consists of two components: the bulge and the
halo. The bulge is described by the Hernquist (1990) potential:
$\Phi(r)=-GM_e(r+a_e)^{-1}$, where $M_e$ the mass of the bulge and
$a_e$ is a measure of its size. For the halo we use the same model as
in the case of spiral galaxies described above. The Hernquist
potential has a simple analytical form and it reproduces the
brightness profile observed typically in ellipticals. We consider two
extreme cases: a large elliptical with $M_{\rm e}=5\times 10^{11}
M_\odot$, and $a_e = 5$ kpc, and a small elliptical with mass $10^3$
times smaller and size $10$ times smaller. The mass and dimension of
the halo are scaled identically as the bulge. The binaries are placed
on circular orbits with a random angular momentum direction in the
bulge with a mass density corresponding to the Hernquist potential
$\rho(r)= (M_e/ 2\pi) a_e r^{-1} (a_e+r)^{-3}$.  Only binaries with
delay times greater than 1 Gyr are considered for ellipticals, since
these galaxies contain only old populations. For starburst galaxies we
use the same potential model as for spirals, although starbursts are
mostly irregular. We consider a large and a small starburst modeled
by the large and small spiral galaxies. Only binaries with delay times
shorter than 1 Gyr are considered since starbursts are young,
star-forming galaxies.

{\underline {\em Afterglow Calculations.}}  We use the mass density
profiles corresponding to the galactic potentials for the model
galaxies described above to predict the luminosity distribution of the
afterglows. The baryonic fraction is assumed to be 20\% for the disk,
and 0.04\% for the bulge and halo (e.g. Fukugita, Hogan \& Peebles
1998).  We compute the expected luminosities of the afterglows in the [2-10] keV energy band
using the standard synchrotron model
(Sari, Piran \& Narayan 1998). The bursts energy is assumed to be
$E=5\times 10^{49}$ erg, given the observed values of $E_\gamma\sim
5\times 10^{48}$ erg or less (e.g. Fox et al. 2005), and for a typical
efficiency of conversion of total energy into $\gamma$-rays
$\eta_\gamma\sim 0.1$.  Other afterglow parameters are drawn from
random distributions within their typical ranges (see Perna \&
Belczynski 2002 for details).

\section{Results}

In what follows we present and discuss our results obtained from
our reference population synthesis model, as described in great detail
in Belczynski et al.\ 2005. Assumptions regarding primordial binary
characteristics and binary evolution treatment are typical of what is
widely used in the literature. Variations of these assumptions 
(for these too see Belczynski et al.\ 2005) do of course lead to
quantitative differences in the results; however our goal in this
study is to concentrate on the robust characteristics of double
compact objects in the context of the recent short GRB observations. 

{\underline {\em Double Compact Object Formation.}}  Double compact
objects form from massive progenitor systems. The more massive primary
star evolves off the main sequence and eventually fills its Roche lobe
initiating the first mass transfer (MT) episode. The MT is dynamically
stable (due to binary components of comparable mass), and leads to
moderate orbital tightening. The primary loses its envelope and forms
a Helium star, which soon afterwords explodes in a Type Ib supernova (SN)
forming the first compact object. Later on, the secondary evolves off
the main sequence, and fills its Roche lobe while on the red giant
branch, initiating a second MT episode. Most commonly, due to the
extreme mass ratio (first compact object: $\sim 1-2 \msun$ for neutron
stars, secondary $\sim 8-15 \msun$) the MT is dynamically unstable and
the system enters a common envelope phase. The envelope of the secondary
is ejected at the cost of orbital energy, and the system separation is
typically reduced by $\sim 1-2$ orders of magnitude. The system
emerges as a close binary consisting of the first compact object and a
naked helium star (the core of the original secondary). At this point
the evolution may follow two qualitatively different paths.

{\em Classical formation channel:} The helium star evolves and never
fills its Roche lobe. The evolution stops at the point where the
helium star explodes in a Type Ib SN forming the second compact
object. Provided that the SN explosion does not disrupt the binary, a
double compact object is formed on a rather wide orbit (e.g.,
Bhattacharya \& van den Heuvel 1991).

{\em New formation channel:} If the helium star has low mass
($\lesssim 3-4 \msun$), it is known to expand at the later stages of
its evolution (e.g., see Belczynski \& Kalogera 2001 for a discussion
and references). Since the binary is rather tight (after the common
envelope phase) the Helium star at some point overfills its Roche lobe
and initiates a third MT episode. For many cases this MT is
dynamically stable (Ivanova et al.\ 2003; Dewi \& Pols 2003) and the
orbital separation decreases further, until the helium star explodes
in a Type Ib/c SN forming the second compact object. A double compact
object is then formed with a very tight (ultracompact) orbit (e.g.,
Belczynski \& Kalogera 2001; Belczynski et al.\ 2002a,c; Ivanova et
al. 2003).

{\underline {\em Merger Times.}}  There are two characteristic times
related to the formation and the subsequent evolution of double
compact object binaries. First, there is an evolutionary time: the
time required for the initial progenitor binary (two components on
Zero Age Main Sequence) to form a binary with two compact
objects. Second, there is a merger time, which is set by the orbital
decay of a double compact object binary due to the emission of
gravitational radiation. The delay time (sum of the evolutionary time
and merger time) as well as merger time distributions for NS-NS and
BH-NS binaries are shown in Figure~\ref{tmerger}. 

It can be seen that merger time distributions are bimodal.  The very
tight binaries ($t_{\rm mer} \sim 0.001-0.1$ Myr) originate from the
new formation channel described above. They involve progenitors which
experience an extra MT episode, that leads to additional orbital decay
and thus the formation of systems with very tight orbits.  Long-lived
binaries (with $t_{\rm mer} \sim 100$ Myr -- 15 Gyr) are formed
through classical channels.  On the other hand, delay time
distributions in a range of 10 Myr -- 15 Gyr are rather flat, with
prominent peaks at $t_{\rm del} \sim 20$ Myr. Evolutionary times of
double compact objects are of the order of 10--20 Myr, therefore the
systems with very short merger times are shifted in the delay time
distribution toward higher values, and in particular they form the
peak around 20 Myr. The flat plateau is created by long-lived
binaries.  
In the model presented here (our ``standard'' model), we have adopted
a maximum NS mass of $2\,\msun$. Compared to our standard model, for a
maximum NS mass of $3\,\msun$, approximately 90\% of the BH-NS become
NS-NS and the remaining BH-NS binaries (10\%) are wide binaries formed
through the classical channels. The highest NS mass estimate is $2.1 
\pm 0.2 \msun$ for a millisecond pulsar in PSR J0751+1807; a relativistic 
binary with helium white dwarf secondary (Nice et. al. 2005).

{\underline {\em Merger Rates.}}  
Throughout the paper, we adopt a flat
cosmology ($\Omega_{\rm m}=0.3$, $\Omega_\Lambda=0.7$ and
$H_0=70$ km s$^{-1}$ Mpc$^{-1}$, and use the star-formation history (corrected for extinction) presented
by Strolger et al.\ (2004). We should however point out that, in general,
the star formation rate is expected to be rather different in ellipticals and in spirals,
since most ellipticals were assembled before $z\sim 2$ and
they are no longer forming stars, while spirals and starbursts galaxies have
an on-going and active star formation (see Smith et al. 2005 for a discussion). 
We defer this whole issue of the study of the relative contribution of the two
types of galaxies to another paper (O'Shaughnessy et al. in preparation).
  
The merger rates of NS-NS and BH-NS binaries as a function of redshift
are shown in Figure~\ref{rates}.  We obtain delay times (time
to formation of double compact object plus merger time) and mass
formation efficiency from population synthesis (for details see Belczynski et al.\
2002b).  The predicted merger rates as a function of redshift are a
convolution of the adopted star formation rate history and the delay
times characteristic for NS-NS and BH-NS mergers: The longer the delay
times, the greater is the shift of merger events to lower
redshifts. With the model distributions of delay times, the peak of
NS-NS mergers appears at redshift $\sim 1$ instead of $\sim 3$ of the
star formation curve (Figure~\ref{rates}). The absolute normalization
of the merger rates in Figure~\ref{rates} is arbitrary, since its
value is subject to large uncertainties (1--2 orders of magnitude)
due to the some poorly constrained population synthesis model
parameters (e.g., Belczynski et al.\ 2002c).

{\underline {\em Merger Locations.}}  We present the distributions of
merger locations for different host galaxies in
Figure~\ref{locations}. In starburst galaxies, the double compact
object population is dominated by new systems with short-merger times, and
therefore most of the mergers are expected to be found within hosts
(more so for massive galaxies). A small fraction (10-30\%, depending
on host mass) of mergers takes place outside hosts and these are mergers
of classical systems with merger times $\sim 10-1000$ Myrs, which are
allowed in the model populations due to our rather long adopted age of
starburst (1 Gyr).  In elliptical galaxies, a substantial fraction of
mergers takes place outside hosts at present. In particular, $\sim$~80\%  
and $\sim$~30\% of NS-NS and BH-NS mergers may take place outside of 
small and large host,  respectively. However, we
note that even for small hosts we find a small but significant
fraction ($\sim$~20-30\%) of mergers within several kpc from the host
center. Spiral galaxies hosting both young and old stellar populations
represent the intermediate case between starburst and ellipticals.

\begin{figure}
\centering
\plotone{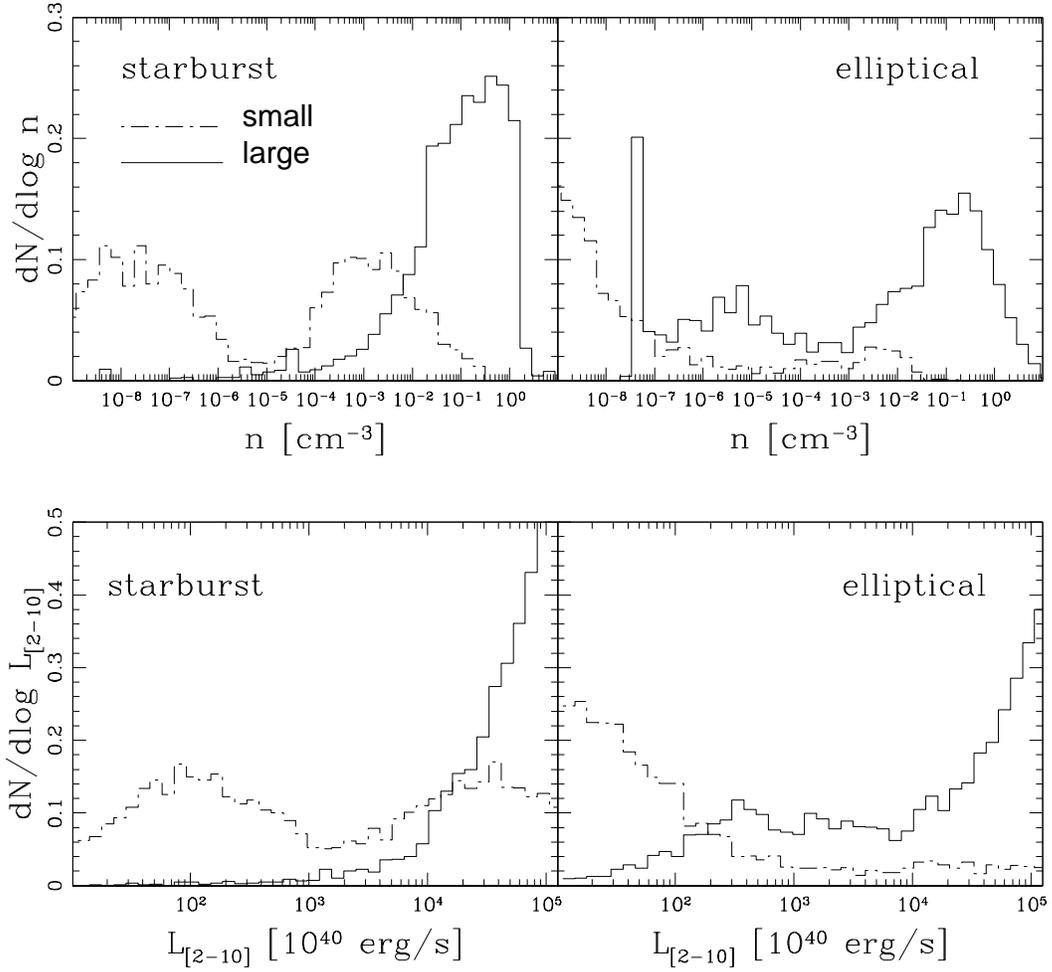}
\caption{Probability distribution for density (top) and 
luminosity in the [2-10] keV band (bottom) for different environments in which
the mergers take place.}
\label{after}
\end{figure}

{\underline {\em Afterglows.}}  Figure~\ref{after} shows the
distribution of the ISM number density in the merger sites, 
and the corresponding afterglow luminosity in the [2-10] keV band, for
the cases of a starburst and an elliptical of small and large
dimension and mass. In ellipticals,
especially small ones, most mergers occur at rather low densities,
therefore resulting in generally dimmer afterglows, a fraction of which
could remain undetectable. Such a case of a "naked" burst
might be GRB 050911 (Page et al. 2005). For massive ellipticals the
majority of mergers take place inside hosts and produces generally
detectable afterglows.  In large starbursts, mergers take place within
the hosts and give rise to rather bright afterglows (due to high
typical ISM densities). In small starbursts, the dominant short-lived
double compact object population merges within hosts and produces
detectable afterglows. However, due to our rather high adopted age of
the starburst (1 Gyr) some systems have longer delay times and some
can escape from their hosts, producing very dim afterglows. These dim
afterglows are not expected for very young starbursts (10--100 Myr).

\section{Discussion}

\begin{figure}
\centering
\plotone{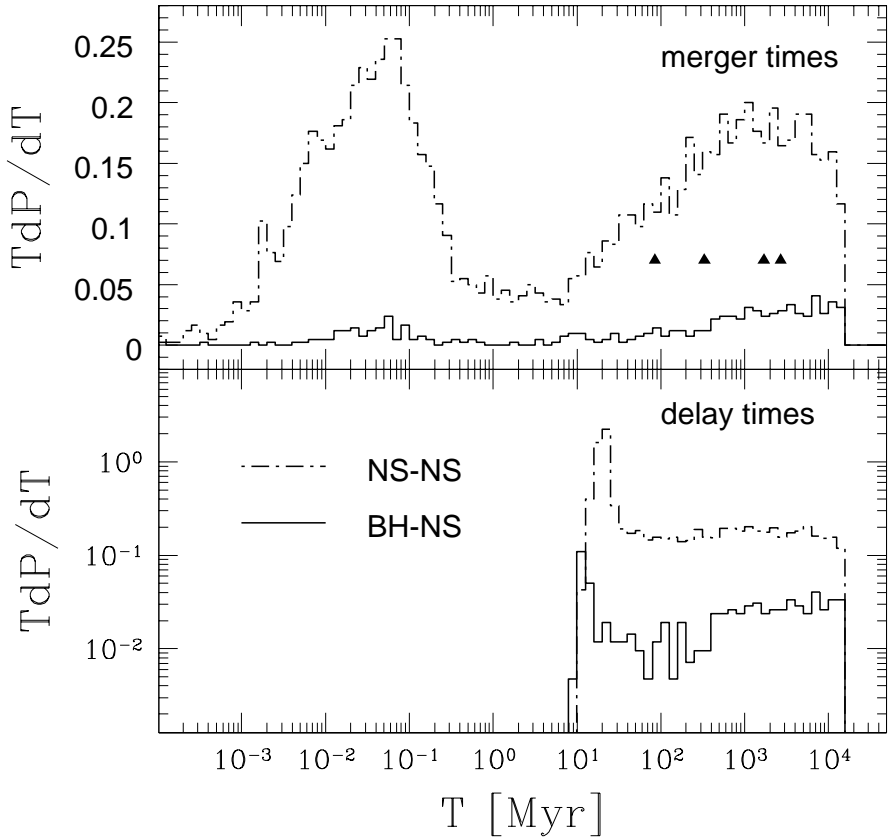}
\caption{ Top: merger time distributions for NS-NS and BH-NS
coalescing binaries. The four field Galactic NS-NS systems are shown
with triangles. Bottom: delay time distributions. Delay time includes
both formation time of a double compact object binary ($\sim 20$ Myr)
as well as its merger time.  Note the different vertical scales on the
panels.}
\label{tmerger}
\end{figure}

\begin{figure}
\centering
\plotone{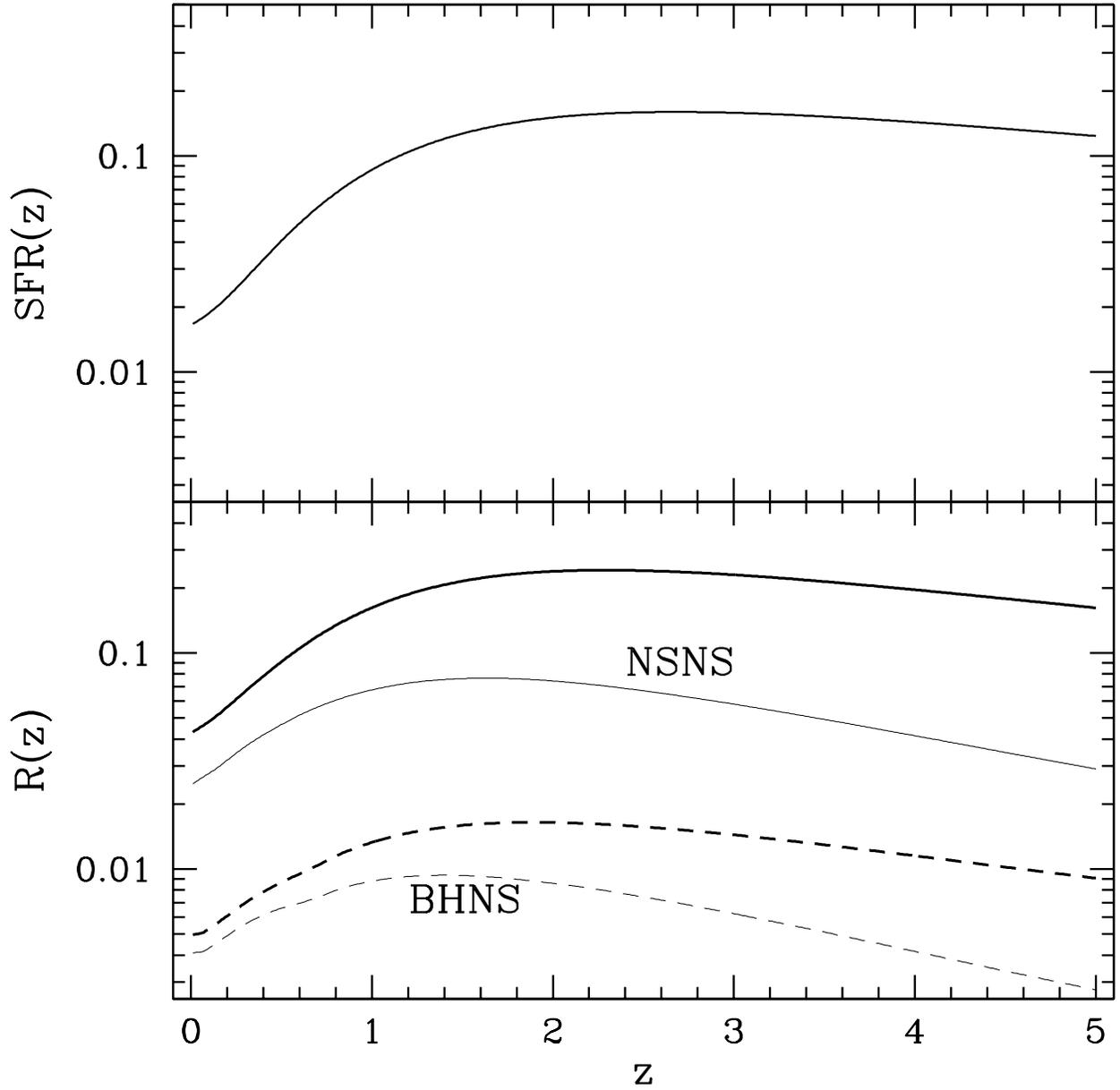}
\caption{The top panel shows the star formation rate history we have used. The bottom panel
(thick lines) shows the inferred merger rate as a function of redshift for the 
NS-NS and BH-NS mergers using the delay times distribution of Figure~\ref{tmerger}.
The thin lines show the contribution from the long-lived ($t>100$ Myr) population. 
Rates are in arbitrary units.}
\label{rates}
\end{figure}

We find two distinct populations of binary compact objects in terms of
their binary-orbit characteristics and associated merger times due to
differences in their evolutionary history. One is a classical
population of rather wide, long-lived systems, with merger times of
$\sim 100-15,000$\,Myr, and the other consists of tight, short-lived
systems with merger times of $\sim 0.001-0.2$\,Myr. In a typical binary 
evolution model, there are roughly similar numbers of short- and long-lived 
NS-NS systems, while there are $\sim$~20\% and $\sim$~80\% of short- and 
long-lived BH-NS systems, respectively. The four known Galactic double 
neutron stars belong to the long-lived classical systems (see Fig.~\ref{tmerger}), 
and, given the small number of NS-NS systems, we do not expect to see the 
short-lived systems, since they merge very soon after their formation.

\begin{figure}
\centering
\plotone{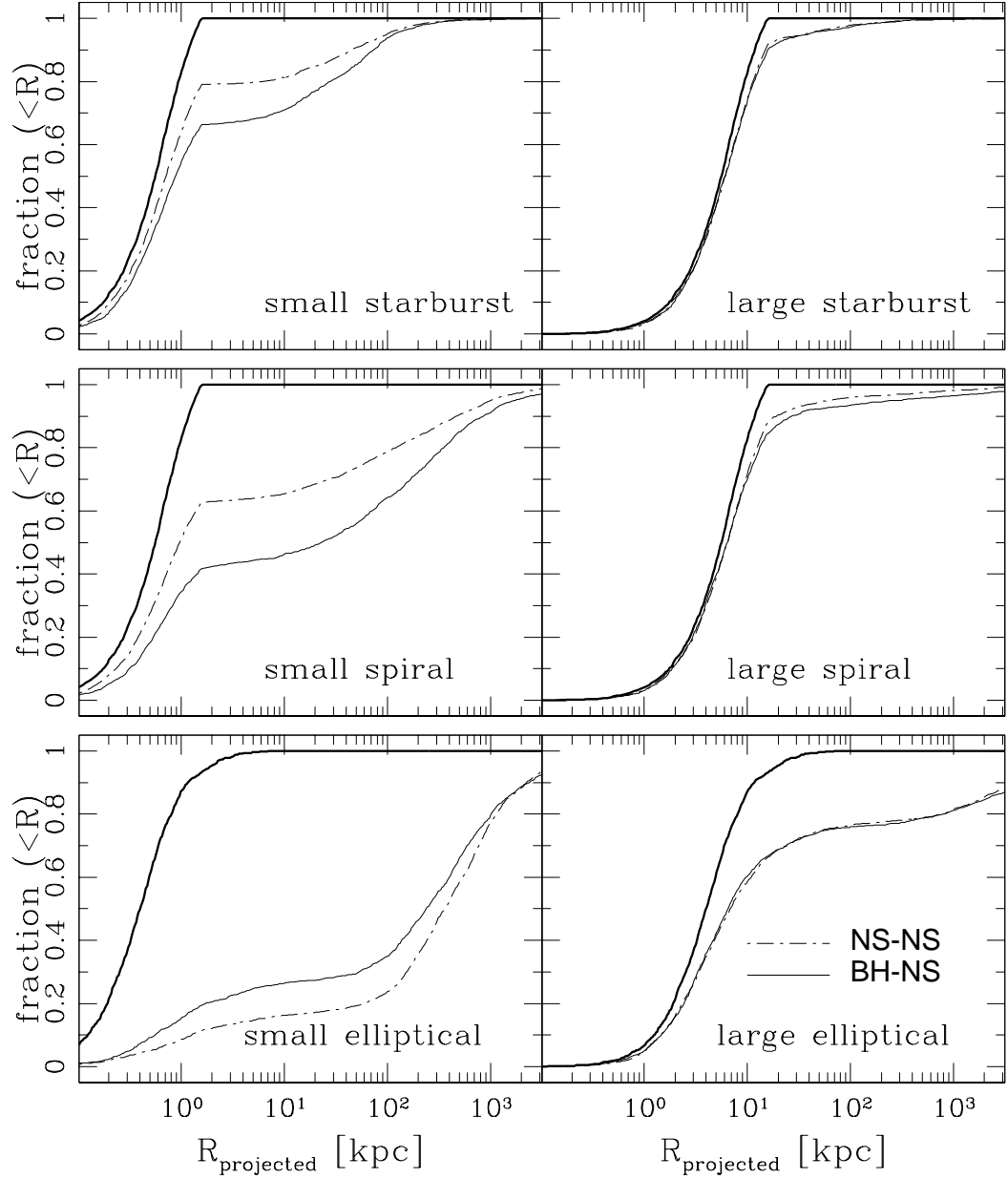}
\caption{ Cumulative distributions of double compact objects merger
locations for different types of host galaxies. Initial distributions
of binaries within each galaxy are shown with thick solid lines. }
\label{locations} \end{figure}

We find that most of the double compact binaries ($\gtrsim$~80\%) that are 
formed in {\em large} galaxies merge within them, independent of the galaxy 
type. For {\em small} galaxies, we find that $\sim$~20\%, 50\%, 70\% mergers 
take place within elliptical, spiral and starburst hosts respectively 
(see Fig.~\ref{locations}). The combination of binary compact 
object lifetimes and their merger locations leads to a testable prediction 
for the location of mergers in relation to the host galaxies. For starburst 
galaxies (which on average have small masses) where stellar populations are 
young, we expect mergers from short-lived double compact objects. These are 
expected to take place
inside or in the vicinity of their hosts. It is worth noting that such
locations, and more importantly, the mere existence of double compact
object mergers associated with star-forming, young galaxies, is not
predicted by any of the previous studies that considered only
classical formation scenarios (long-delayed mergers). If it is shown
that the star-forming host galaxy of GRB~050709 does not contain a
dominant old underlying stellar population, this association will
stand as the ``smoking gun'' evidence for the new short-lived double
compact object formation channel, originally identified by Belczynski
\& Kalogera (2001) and Belczynski et al.\ (2002a,c). 

In elliptical galaxies, which have little or no ongoing star formation, we 
expect to find mergers of long-lived double compact objects at present. For 
massive hosts, most mergers should occur within the hosts, while for smaller 
galaxies outside the hosts. GRB~050509B was tentatively associated with a 
large elliptical galaxy, and its error circle indicates that this burst took 
place either within or in the outskirts of the host, in agreement with our 
findings.  GRB~050911 appears to be a case of a ``naked'' burst (Page et 
al.\ 2005).  It could possibly be explained by our models with a double 
compact object binary originating from a small elliptical galaxy and merging 
far outside the host without any detectable afterglow. On the other hand 
GRB~050724 was found inside a small elliptical.  Our models are consistent 
with this observation, since $\sim$~20\% of mergers are predicted to occur 
within 5 kpc of the center for small ellipticals. 

However, if more short GRBs are found at such intra-galactic
locations, a number of new possibilities may be favored: BH-NS mergers
with small or zero BH kicks, or in general small kicks for both NS and
BH (e.g., Podsiadlowski et al.\ 2004; Dewi et al.\ 2005). Such
alternative models would then have to be investigated.  We note that,
at present, the necessity of imparting such small kicks to the
majority of NS is an open question (c.f., Willems et al.\ 2004;
Chaurasia \& Bailes 2005; Ihm et al.\ 2005).  This, combined with our
results for the locations of the mergers, indicates that short GRBs in
the outskirts of galaxies need not be associated with globular
clusters, as recently claimed by Grindlay et al.\ (2006).

We find that short GRB progenitors originate most probably from a
diverse population of compact objects which are formed in old but also
in young stellar environments. Although the absolute formation rates
and redshift distributions implied by a mixture of galaxy types must be 
investigated and compared with theoretical predictions in
more detail (O'Shaughnessy et al., in preparation), at present we are
able to account for the origin (and associated delay times) of several
recently observed short GRBs. 

Some of the recent work (e.g., Nakar, Gal-Yam \& Fox 2005) stands in 
apparent contradiction with our findings, presenting claims that the 
current  models of double compact objects cannot explain the new short 
GRB observations. Also, long delay times ($\sim 6$ Gyr) are inferred 
from the associations of short GRBs with old elliptical galaxies (Nakar 
et al.\ 2005). We note that at present these should be considered with 
caution. First, they are based on the analysis of a very small number of 
bursts and may suffer from small number statistics; a significant fraction 
of short delay bursts cannot be ruled out with high statistical 
significance. The comparison in Nakar et al.\ (2005) has been performed 
using some trial functions of delays, which do not necessarily correspond 
to the actual delays obtained in detailed population synthesis calculations. 
Second, their analysis neglects the information about the types of 
individual short GRB host galaxies and their particular star formation 
histories.  
Third, the models of Nakar et al.\ (2005) cannot explain the observed Galactic 
population of relativistic double neutron stars, with merger times of $\sim
100-3000$ Myr, that are consistent with our models (see Fig.~\ref{tmerger}).
Finally, the estimate of the delays may be complicated by 
evolutionary effects. We do know that the population of long GRBs is 
affected by cosmological evolutionary effects, for example, related with the 
metallicity evolution of the Universe, and so may be the population of short 
GRBs. Thus a direct comparison with the total star formation rate as performed 
by Nakar et al.\ (2005) may be misleading. The influence of such evolutionary 
effects and of the contribution of different types of galaxies and their star 
formation history will be examined in detail in O'Shaughnessy et al.\ (2006). 

Additionally, the findings of Nakar et al.\ (2005) should be weighted by the 
observations of GRB~050724 and GRB~051221A in young star forming environment, 
as well as the detection of GRB~050813 at z=1.7-1.9 where there ought be no 
short GRBs, if the delays are as long as they claim. If more short GRBs are 
found at high redshifts ($z \gtrsim 1$), this will provide further support 
for delay times, that at least for some short GRBs, are shorter than $\sim 6$ 
Gyr. It has also been suggested that there is an observational bias against 
detecting short GRBs at high redshifts with {\em SWIFT}; the minimum flux for 
detection is apparently smaller in the observed sample than for GRBs with secure 
redshift determination, and therefore it favors redshift determinations for 
closest (low redshift) bursts (Hopman et al.\ 2006). If this is the case, then 
it is clear that the observed redshift distribution leads to an overestimate in 
delay times in the analysis of Nakar et al.\ (2005). 

We conclude that the current observations point toward a short-hard GRB 
progenitor population that is diverse in terms of merger times and locations, 
as suggested by our models. Soon, with more observations of short GRBs, we will 
be able to critically test the population synthesis models. Specifically, we 
will investigate effects of low metallicity, binary populations dominated by 
roughly equal-mass binaries as suggested by Pinsonneault \& Stanek (2006) and 
low natal kicks proposed by Pfahl et al.\ (2002) on progenitors of double neutron 
star systems, and their relevance for both short GRB and gravitational-wave 
observations (Belczynski et al.\ 2006, in preparation).

\acknowledgments This work was partially supported by KBN grant
1P03D02228 (KB, TB); NASA grant NNG05GH55G and NSF grant AST~0507571 (RP);  
NSF grant PHY-0353111 and a Packard Fellowship in Science and Engineering (VK); 
NSF grant PHY 99-0794 (NI); and NASA contract NASW-4690 and grant NAGW5-10579 (DQL).

\clearpage

\end{document}